\documentclass{article}

\usepackage{PRIMEarxiv}

\usepackage[utf8]{inputenc} 
\usepackage[T1]{fontenc}    
\usepackage{hyperref}       
\usepackage{url}            
\usepackage{booktabs}       
\usepackage{amsfonts}       
\usepackage{nicefrac}       
\usepackage{microtype}      
\usepackage{lipsum}
\usepackage{fancyhdr}       
\usepackage{graphicx}       
\graphicspath{{media/}}     

\usepackage{amsthm} 
\usepackage{amsmath} 
\usepackage{booktabs}  
\usepackage{threeparttable}
\usepackage{multirow}
\usepackage{comment}
\usepackage{bm}
\usepackage{subcaption}
\usepackage{algpseudocode}
\usepackage[ruled,linesnumbered]{algorithm2e}
 
\newtheorem{definition}{Definition}[section] 
\newtheorem{theorem}{Theorem}

\title{PerCNet: Periodic Complete Representation for Crystal Graphs
\thanks{Corresponding authors: Qianli Xing (qianlixing@jlu.edu.cn) and Bo Yang (ybo@jlu.edu.cn)}
}
\author{
  Jiao Huang \\
  Jilin University \\
  Changchun, Jilin \\
  \texttt{haungjiao20@mails.jlu.edu.cn} \\
  \And
  Qianli Xing $^{*}$\\
  Jilin University \\
  Changchun, Jilin \\
  \texttt{qianlixing@jlu.edu.cn} \\
  \And
  Jinglong Ji \\
  Jilin University \\
  Changchun, Jilin \\
  \texttt{jijl22@mails.jlu.edu.cn} \\
   \And
  Bo Yang $^{*}$\\
  Jilin University \\
  Changchun, Jilin \\
  \texttt{ybo@jlu.edu.cn} \\
}

\begin{document}
\maketitle

\begin{abstract}
Crystal material representation is the foundation of crystal material research. 
Existing works consider crystal molecules as graph data with different representation methods and leverage the advantages of techniques in graph learning.
A reasonable crystal representation method should capture the local and global information. 
However, existing methods only consider the local information of crystal molecules by modeling the bond distance and bond angle of first-order neighbors of atoms, which leads to the issue that different crystals will have the same representation. To solve this many-to-one issue, we consider the global information by further considering dihedral angles, which can guarantee that the proposed representation corresponds one-to-one with the crystal material.  
We first propose a periodic complete representation and calculation algorithm for infinite extended crystal materials.
A theoretical proof for the representation that satisfies the periodic completeness is provided. 
Based on the proposed representation, we then propose a network for predicting crystal material properties, PerCNet, with a specially designed message passing mechanism. 
Extensive experiments are conducted on two real-world material benchmark datasets. The PerCNet achieves the best performance among baseline methods in terms of MAE. In addition, our results demonstrate the importance of the periodic scheme and completeness for crystal representation learning.
\end{abstract}


\section{Introduction}

Crystal material representation is the foundation of many key domains including crystal molecule property prediction \cite{xie2018crystal, chen2019graph, choudhary2021atomistic, yan2022periodic, lin2023efficient} and crystal material generation \cite{zhao2023physics, xie2021crystal}. 
A reasonable crystal representation should fully capture the crystal's global information, such as its crystal structure, as well as local information, including atomic characteristics and interactions between atoms. 
Thus, a reasonable crystal representation can enhance the predictive accuracy for property prediction tasks and the quality of generated crystals for crystal structure generation tasks.

\begin{figure}[ht!]
    \centering
    \includegraphics[clip, trim=0 5.5cm 0 5cm, scale=0.3]{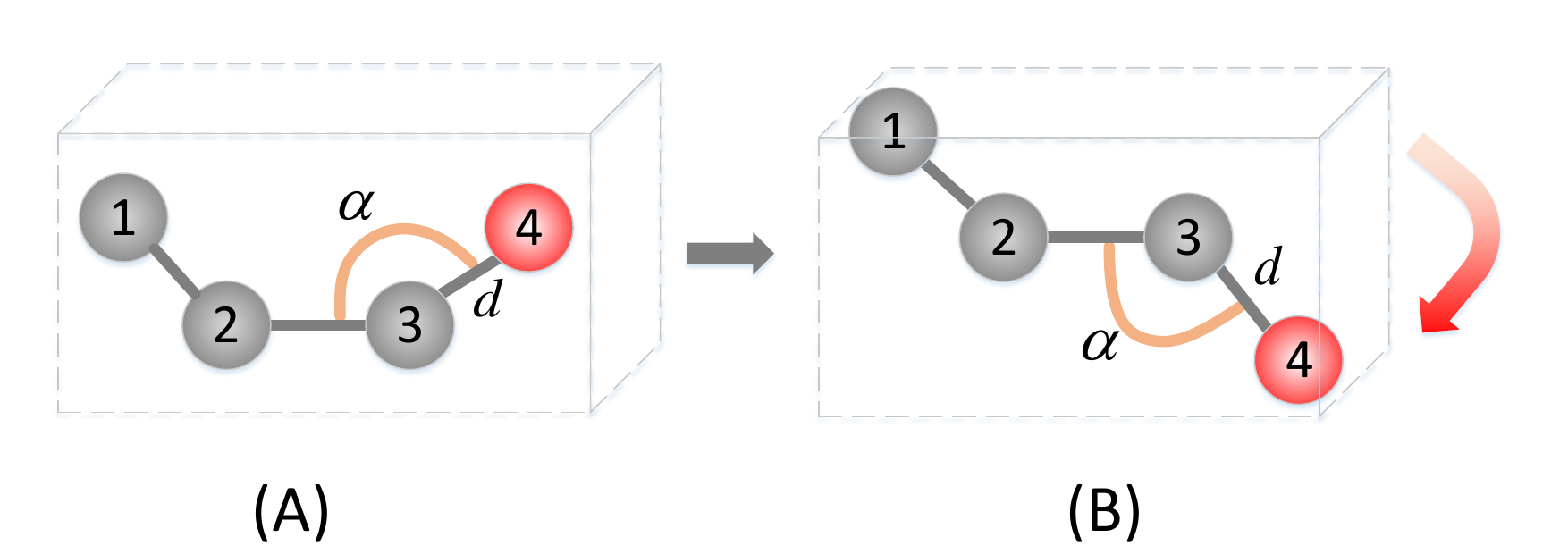}
    \caption{Illustrations of cases that existing crystal representation methods cannot distinguish. 
    (A). An illustration of a crystal molecule with four atoms within the unit cell. 
    (B). An illustration of a crystal structure similar to (A) but with minor differences.
    The new structure changes the coordinate position of atom 4 while keeping the distance such as $d$ and the angle such as $\alpha$ between atom 4 and its first-order neighbors unchanged. 
    }
    \label{fig:advantage}
\end{figure}

To obtain crystal representation, researchers attempt to employ the organic molecule representation methods, which already have a variety of achievements in fields of molecular property prediction \cite{wang2022advanced, borgwardt2005protein}, protein interface contact prediction \cite{fout2017protein, liu2020deep, morehead2021geometric}, proteins generation \cite{jumper2021highly, wang2023multitask} and so on.
The main reason is that crystal molecules and organic molecules have similar local information such as atomic features and the modes of interaction between atoms.
Nevertheless, crystal molecules and organic molecules differ significantly in their overall macroscopic structures. Organic molecules are composed of a finite number of atoms, while crystals consist of an infinite array of repeating units, each containing a finite number of atoms. Hence, for crystals, mastering the intricacies of this infinitely extending pattern poses a significant challenge.

In order to address it, CGCNN \cite{xie2018crystal} first represents crystal molecular structure as an undirected multigraph where nodes represent atoms in the unit cell and edges represent atom connections. The periodicity scheme is modeled by multiple edges between atoms, and the effectiveness of this modeling method is verified on crystal property prediction tasks.
Subsequent works \cite{gasteiger2021gemnet, louis2020graph, choudhary2021atomistic, zhao2023physics, chen2019graph, yan2022periodic} are mostly improving their representation based on this scheme.
Specially, Matformer formally proposes the concept of periodic invariance, further emphasizing the importance of periodicity.
By infinitely summing distances, the recently released PotNet \cite{lin2023efficient} proposes a competitive crystal molecule representation method that models the complete set of potentials among all atoms and achieves state-of-the-art (SOTA) performance in crystal property prediction tasks.
Despite some processes in crystal materials representation, current work only considers the local information that can not guarantee that the representation corresponds one-to-one with the crystal material. 
For example, as shown in Figure \ref{fig:advantage}, current crystal representation works can not distinguish these two structures because they only consider the bond distance and bond angle of first-order neighbors. 
As a result, different kinds of crystals may have the same representation (many-to-one issue), constraining the performance ceiling of the model. This subsequently leads to lower predictive accuracy in tasks related to material property predictions, or lower quality in tasks involving material structure generation.

To deal with the above many-to-one issue, we model the global structure information by further considering the dihedral angles of first-order neighbors which can guarantee the proposed representation corresponds one-to-one with the crystal material. 
We first propose a periodic complete representation for infinite extended crystal materials. 
To verify the effectiveness of our proposed representation, we provide theoretical proof of the periodic completeness and further show in real experiments that the proposed representation method can distinguish the crystal materials (e.g. the example in Figure \ref{fig:advantage}), while others can not.
Furthermore, we propose a network for predicting crystal material properties, PerCNet.
We test the performance of PerCNet on two publicly available large-scale datasets including The Materials Project and JARVIS, and the final results show that PerCNet outperforms baseline methods.

Our contribution mainly includes the following three parts:
\begin{itemize}
\item To our best known, PerCNet is the first work that ensures the representation corresponds one-to-one with the crystal material.

\item The theoretical proof for the representation corresponding one-to-one with the crystal material is provided. Furthermore, real experiments are conducted to verify the effectiveness of the proposed representation.

\item PerCNet achieves the best performance among baseline methods for property prediction tasks in two real-world material benchmark datasets \cite{choudhary2020joint, jain2013materials}.
\end{itemize}

\section{Related work}

Over the past decade, machine learning has seen a surge in developments for organic molecule studies, such as small molecules \cite{duvenaud2015convolutional,wang2022advanced,wu2018moleculenet} and proteins \cite{borgwardt2005protein,fout2017protein,jumper2021highly,liu2020deep,morehead2021geometric}.
More recently, the advent of graph neural networks (GNNs) has introduced an innovative approach by representing molecular structures as 2D graphs, where atoms are represented as nodes and chemical bonds as edges \cite{gao2019graph, gao2021topology, liu2021dig, xu2018powerful, zhang2018end}. 
However, this method fails to fully capture the structure of real-world 3D molecules due to the loss of information during the 2D transformation process. This has given rise to the need for 3D graph neural networks (GNNs) \cite{gasteiger2020fast, shuaibi2021rotation, gilmer2017neural, sanchez2020learning, vignac2020building, battaglia2018relational}. 
In these methods, atoms are mapped from three-dimensional Cartesian coordinates to relative variables like distance and angle to satisfy SE(3)-invariance requirements.
Notably, SphereNet \cite{liu2021spherical} and ComENet \cite{wang2022comenet} have proposed comprehensive molecular representations for 3D graphs. 
However, these representation methods for molecules of limited size fail to model periodicity, making them unsuitable for representing crystal molecules. 

Encouraged by the success of GNNs on organic molecules, current research \cite{ramprasad2017machine, meredig2014combinatorial, oliynyk2016high, raccuglia2016machine, ward2016general} has shown significant interest in crystal materials. 
CGCNN \cite{xie2018crystal} was the first to propose a multi-graph representation for materials, allowing multiple edges between the same pair of end nodes to encode both atomic information and bonding interactions between atoms, a departure from typical molecular graphs. 
Building on this, Matformer \cite{yan2022periodic} proposed a new architecture with periodic invariance for crystals. 
Additionally, PotNet \cite{lin2023efficient} modeled the complete set of potentials among all atoms by summing distances infinitely.
However, current work only focuses on local information such as bond length and bond angle, which results in their representations not being able to correspond one-to-one with crystal molecules.

\section{Notations and Definations}

\subsection{Crystal Notations and Property Prediction}

\begin{figure}[t]
    \centering
    \includegraphics[clip, trim=0 3cm 0 4cm, scale=0.3]{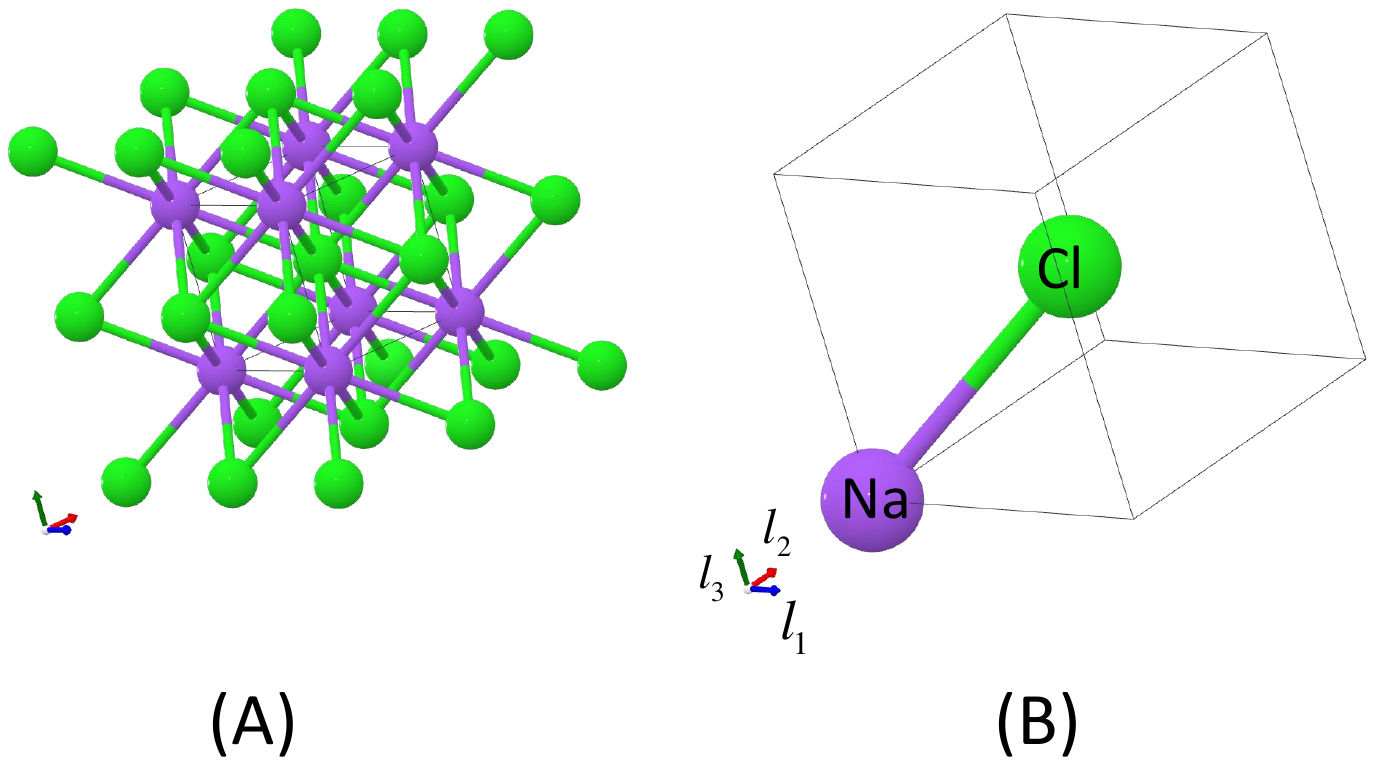} 
    \caption{Illustrations of the crystal molecular structures of NaCl, which is identified by its Material Project \cite{jain2013materials} ID, mp-22851.
    (A). An illustration of the infinite periodic crystal structure of NaCl. (B). An illustration of a unit cell and Lattice of Nacl. The unit cell is the region inside the parallelepiped, containing both Na and Cl atoms. The lattice is represented by the three arrows at the bottom left corner, guiding the periodic expansion direction of the unit cell. The coordinates of atoms and lattices are detailed in Table \ref{tab:unitcell}. }
    \label{fig:unitcell}
\end{figure}              

\begin{table}[t]
    \centering
    \caption{Coordinates of atoms and lattices for NaCl.}
    \begin{tabular}{|c|c|c|c|c|}
        \hline
        \multicolumn{2}{|c|}{Coordinate}&x&y&z\cr
        \hline
        \multirow{2}{*}{Unit Cell}
         & Na & 0 & 0  & 0 \cr
         \cline{2-5}
         & Cl & 1.751 & 1.751& 1.751 \cr
         \hline
         \multirow{3}{*}{Lattice} 
         & l\(_1\) & 3.502 & 0& 0 \cr
         \cline{2-5}
         & l\(_2\) & 0 & 3.502& 0 \cr
         \cline{2-5}
        & l\(_3\) & 0& 0 & 3.502 \cr
        \hline
    \end{tabular}
    \label{tab:unitcell}
\end{table}
To start, we'll establish the notations used throughout this paper.
A crystal material's structure can be depicted as the infinite extension of its unit cell, with the unit cell serving as the smallest representative entity illustrating the crystal's structure.
As exemplified in Figure \ref{fig:unitcell}, we showcase the molecular structure of NaCl (mp-22851). A crystal structure can be expressed as $\bm{M} = \{\bm G, \bm L\}$, where $\bm G = \{\bm A, \bm X \}$ describes the unit cell consisting of N atoms. 

Here, $\bm A = [a_1,..., a_N]^\top \in \mathbb{A}^N$ denotes atom types, with $A$ representing the set of chemical elements. For NaCl (mp-22851), we define A as $A=\{Na, Cl\}$.
Moreover, $\bm X = [\bm x_1, ..., \bm x_N]^\top \in \mathbb{R}^{N\times3}$ specifies the three-dimensional coordinates of atoms in the Cartesian coordinate system. 
$\bm L = [\bm l_1, \bm l_2, \bm l_3]^\top \in \mathbb{R}^{3\times3}$ represents the periodic lattice, indicating the directions in which the unit cell extends infinitely in three-dimensional space. The values of $\bm X$ and $\bm L$ for NaCl (mp-22851) are displayed in Table \ref{tab:unitcell}.

Formally, given a crystal structure $\bm M = \{\bm G, \bm L\}$, with $\bm G = \{\bm A, \bm X\}$, the infinite crystal structure can be represented as:
\begin{equation}
    \bm{M} = \bm G^{all} = \{(a_i^{all}, \bm x_i^{all})|a_i^{all}=a_i,
    \bm x_i^{all} = \bm x_i+\bm K\odot \bm L, 
    \bm K= [k_1, k_2, k_3]\in \mathbb{Z}^{3}\}.
\end{equation}

Here, $\bm K\odot \bm L = k_1\bm l_1+k_2\bm l_2+k_3\bm l_3$ clarifies the coordinate transformation of the \textit{i}th atom within the unit cell as it undergoes periodic extension in space.

In this work, our focus lies on the task of crystal property prediction. Given a crystal structure $M=\{A, X, L\}$, the goal is to predict a crystal property value y, which is modeled as $M\rightarrow y\in \mathbb{R}$ for regression tasks. In the domain of crystal molecules \cite{xie2018crystal, chen2019graph, choudhary2021atomistic, yan2022periodic, lin2023efficient}, the predominant focus is on properties that gauge structural stability, such as energy and band gap. These properties will also be the focal points in our subsequent experiments.

\subsection{Definition of Periodic Completeness}
\begin{definition}( Periodic Completeness ).
    For two crystal graph $M_1=\{A, X_1, L_1\}$ and $M_2 =\{A, X_2,L_2\}$ with same atom type $A$, their geometric representations $P$ are termed periodic complete if and only if they satisfy
\begin{equation}
    P_{M_1} = P_{M_2} \Leftrightarrow
    \exists (R_1 \in SE(3), R_2\in E(3)), X_1 = R_1(X_2) \wedge L_1 = R_2(L_2) \label{complete}.
\end{equation}
\end{definition}

Here, SE(3) denotes the special Euclidean group in three dimensions, which includes all rotations and translations in 3D. E(3) represents the Euclidean group in three dimensions, including all rotations, translations, and mirror reflections in 3D.

The formula on the left asserts that the representations of two crystal molecules are identical. The right side of the formula indicates that the coordinates of the two crystal molecules belong to the same SE(3) group, and the lattice coordinates belong to the same E(3) group. In other words, they represent the same crystal molecule.

There are two scenarios where a crystal molecule's representation might fail to satisfy periodic completeness.

Firstly, the one-to-many problem arises when a single crystal material can have multiple representations if the representation does not satisfy invariances. 
For crystal materials, a comprehensive representation must satisfy four types of invariances \cite{xie2021crystal}, including permutation invariance, translation invariance, rotation invariance, and periodic invariance. Existing crystal representation methods, to satisfy spatial isotropy invariance (SE(3) invariance) and Euclidean invariance (E(3) invariance), typically convert all absolute variables, such as three-dimensional Cartesian coordinates, into corresponding relative variables, such as intermolecular distances and angles, to be used as inputs. Our approach also follows this practice.

Secondly, the many-to-one problem occurs when multiple crystal molecules share the same representation. For example, diamond and graphite both have the same molecular formula representation, C, but their properties differ due to their distinct geometric structures. Therefore, it is crucial for the representation method to accurately differentiate between two different geometric structures. However, existing crystal molecular representation methods, which only model interatomic distances and bond angles, struggle to effectively distinguish between similar structures, as shown in Figure \ref{fig:advantage}. To address this issue, we propose a periodic complete representation that can differentiate all distinct crystal structures.

\subsection{Periodic Completeness Representation}

\begin{figure}[ht]
    \centering
    \includegraphics[scale=0.28]{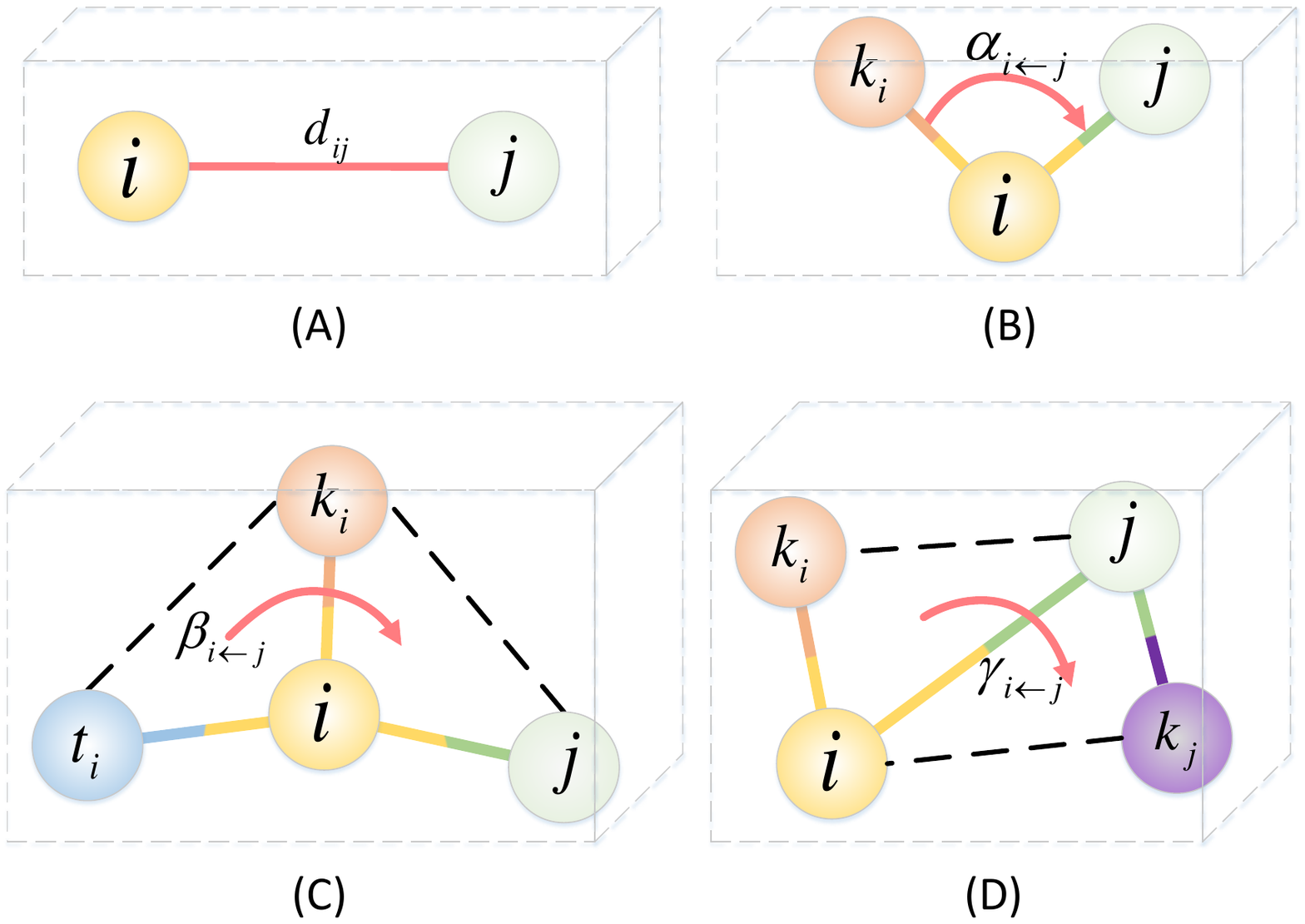}
    \caption{Illustrations of geometric representation between atom i and its neighbor atom j, $P_{i\leftarrow j} = \{ d_{i j}, \alpha_{i\leftarrow j}, \beta_{i\leftarrow j}, \gamma_{i\leftarrow j} \}$.
    (A). An illustration of the distance $d_{i j}$ between atom i and atom j.
    (B). An illustration of the angle $\alpha_{i\leftarrow j}$ between edge $d_{i j}$ and edge $d_{i k_i}$ where atom $k_i$ is the nearest neighbor for atom i.
    (C). An illustration of $\beta_{i\leftarrow j}$, which is defined as the dihedral angle between half plane $ik_it_i$ and half plane $ik_ij$.
    (D). An illustration of $\gamma_{i\leftarrow j}$, which is defined as the dihedral angle between half plane $ijk_i$ and half plane $ijk_j$. Atom $k_j$ is the nearest neighbor for atom j.
    }
    \label{angle}
\end{figure}

\begin{definition}( Periodic Complete Representation ).

\label{def:PMC}
For a crystal material M, its periodic complete representation is given by:
\[
P_M = [(d_{i j}, \alpha_{i\leftarrow j}, \beta_{i\leftarrow j}, \gamma_{i \leftarrow j})]_{i\in I;j\in NB_i}\in \mathbb{R}^{n_{edge}\times4}.
\]
\end{definition}

From Definition \ref{def:PMC}, it can be observed that periodic completeness is achieved by taking into account both local information, such as distance $d$ and bond angle $\alpha$, as well as global information, which includes dihedral angles represented as $\beta$ and $\gamma$.

In this definition, $I$ symbolizes the set of atom indices, ranging from 1 to $N$, where $N$ is the total count of atoms in the unit cell. $NB_i$ signifies the set of all neighboring atoms for a given atom $i$, and $m$ stands for the total number of neighbors for atom $i$. Atoms situated at a distance less than the cutoff value from atom $i$ are considered its neighbors, and the total count of neighbors is represented by $n_{edge}$.

Importantly, the cutoff value is determined to exceed the maximum of $\{max\_con$, $min_{l_1}$, $min_{l_2}$, $min_{l_3}\}$. The term $min_{l_s}$ is employed to define the minimum distance between any pair of atoms, with one atom positioned within the unit cell and the other in the adjacent unit cell along the $\bm l_s$ direction. The term $max\_con$ signifies the shortest edge length, ensuring the graph composed of atoms within the crystal cell is connected. The cutoff value plays a crucial role, as it guarantees the connectivity of all atoms within the crystal cell and the graph formed by their neighbors. It also ensures the complete modeling of the periodic information $\bm L$, a point we will further illustrate in subsequent proofs.

The definition of $\{d, \alpha, \beta, \gamma\}$ is depicted in Figure \ref{angle}. 
$d_{i j}$ represents the edge distance between atom $i$ and atom $j$, where atom $j$ can be any neighbor of atom $i$.
$\alpha_{i\leftarrow j}$ denotes the angle between the edge $d_{i j}$ and the edge $d_{i k_i}$, where atom $k_i$ is the nearest neighbor of atom $i$.
$\beta_{i\leftarrow j}$ is defined as the angle by which the half-plane $ik_it_i$, bounded by the straight line along vector $\overrightarrow{ik_i}$, rotates counterclockwise to overlap the half-plane $ik_ij$, also bounded by the line along vector $\overrightarrow{ik_i}$. Here, atom $t_i$ is the neighbor of atom $i$, and the angle between vectors $\overrightarrow{it_i}$ and $\overrightarrow{ik_i}$ is the smallest angle that is greater than 0 and less than $\pi$. The choice of atoms $k_i$ and $t_i$ is purposeful, as it will be instrumental in the proof of periodic completeness.
Lastly, $\gamma_{i\leftarrow j}$ represents the angle by which the half-plane $ik_jj$, bounded by the straight line along vector $\overrightarrow{ji}$, rotates counterclockwise to overlap the half-plane $ik_ij$, also bounded by the line along vector $\overrightarrow{ji}$. Here, atom $k_j$ is the nearest neighbor of atom $j$.

\begin{algorithm}
\caption{Getting Periodic Complete Representation $P=\{d, \alpha, \beta, \gamma\}$.}
\label{alg1}
\LinesNumbered 
\KwIn{$\bm X = [\bm x_1, ..., \bm x_N]^\top \in \mathbb{R}^{N\times3}; \bm L = [\bm l_1, \bm l_2, \bm l_3]^\top \in \mathbb{R}^{3\times3}$} 
\KwOut{$P_M = [(d_{i j}, \alpha_{i\leftarrow j}, \beta_{i\leftarrow j}, \gamma_{i \leftarrow j})]_{i\in I;j\in NB_i}\in \mathbb{R}^{n_{edge}\times4}$} 

\For{$i=1$ to N}
{
  \For{$j=1$ to $m_i$}
  {
    Compute $d_{i j}$ via: $d_{i j} = ||\mathbf{x_j} - \mathbf{x_i}||_2$ \\
    }
    Get nearest atom $k_i$ for atom $i$ via: $k_i = argmin_{k\in NB_i}(d_{i k})$\\
    Get vector $\overrightarrow{ik_i}$ via: $\overrightarrow{ik_i} = x_{k_i} - x_i$ \\
  \For{$j=1$ to $m_i$}
  {
    Get vector $\vec{ij}$ via: $\vec{ij} = x_j - x_i$\\
    Compute $\alpha_{i\leftarrow j}$ via: $\alpha_{i\leftarrow j} = angle(\overrightarrow{ik_i}, \vec{ij})$\\
  }
  Get atom $t_i$ via: $t_i = argmin_{t\in NB_i}( 0<\forall \alpha_{i\leftarrow t}<\phi )$\\
  \For{$j=1$ to $m_i$}
  {
    Get vector $\vec{it_i}$ via: $\overrightarrow{it_i} = x_{t_i} - x_i$\\
    Get normal vector $\vec{ik_it_i}$ of half plane $ik_it_i$ via vector $\overrightarrow{ik_i}$ and vector $\overrightarrow{it_i}$\\
    Get normal vector $\vec{ik_ij}$ of half plane $ik_ij$ via vector $\overrightarrow{ik_i}$ and vector $\overrightarrow{ij}$\\
    Compute $\beta_{i\leftarrow j}$ via: $\beta_{i\leftarrow j} = angle(\overrightarrow{ik_it_i}, \overrightarrow{ik_ij})$\\
  }
}
\For{$i=1$ to N}
{
  \For{$j=1$ to $m_i$}
  {
    Get vector $\overrightarrow{jk_j}$ via: $\overrightarrow{jk_j} = x_{k_j} - x_j$\\
    Get normal vector $\vec{ik_jj}$ of half plane $ik_jj$ via vector $\vec{ij}$ and vector $\overrightarrow{jk_j}$\\
    Get normal vector $\vec{ik_ij}$ of half plane $ik_ij$ via vector $\vec{ij}$ and vector $\overrightarrow{ik_i}$\\
    Compute $\gamma_{i\leftarrow j}$ via: $\gamma_{i\leftarrow j} = angle(\overrightarrow{ik_jj}, \overrightarrow{ik_ij})$\\
  }
}
\end{algorithm}

The process of obtaining the periodic complete representation $P = \{d, \alpha, \beta, \gamma\}$ from the original geometric information $\{\bm X, \bm L\}$, represented as Cartesian coordinates, is outlined in Algorithm \ref{alg1}.

First, the distance $d$ between atoms is directly calculated using the formula in the third row.

 Following this, lines 5-9 of the algorithm illustrate the calculation of the angle $\alpha$. 

The calculation process for the dihedral angle $\beta$, which requires the computation of two normal vectors, is then demonstrated in lines 11-16. It's important to note that the choice of atom $t_i$ is critical to ensuring that vector $\overrightarrow{it_i}$ and vector $\overrightarrow{ik_i}$ are not collinear, which guarantees that these two vectors can form a plane. This, in turn, ensures the uniqueness of the normal vector $\vec{ik_it_i}$.
Subsequently, a calculation process similar to that used for $\beta$ is employed to obtain $\gamma$.

Regarding computational complexity, the first loop has a complexity of $O(N)$, where N is the number of atoms within the unit cell. 
The second loop has a complexity of $O(M)$, where M is the average number of neighbors for each atom. Each formula inside the loop has a computational complexity of $O(1)$. 
Consequently, the overall computational complexity of the algorithm for obtaining representation $P=\{d, \alpha, \beta, \gamma\}$ is $O(NM)$, which is on par with the time complexity of algorithms that consider only distance or angle.

The ability to incorporate the dihedral angle without increasing time complexity stems from the fact that the calculation of $\{d_{i j}, \alpha_{i\leftarrow j}, \beta_{i\leftarrow j}, \gamma_{i \leftarrow j}\}$ relies solely on the first-order neighbor information of atoms $i$ and $j$, where $k_i, t_i$ are neighbors of atom $i$ and $k_j$ is a neighbor of atom $j$.
In subsequent experiments, we conduct a quantitative analysis of the time consumption of the entire algorithm architecture, further demonstrating the efficiency of our algorithm.

\subsection{Proof of Geometric Periodic Completeness}\label{proof}
\begin{figure*}[ht]
  \centering
  \begin{subfigure}[b]{0.5\textwidth}
    \includegraphics[clip, trim=6cm 10cm 2cm 3.5cm, scale=0.38]{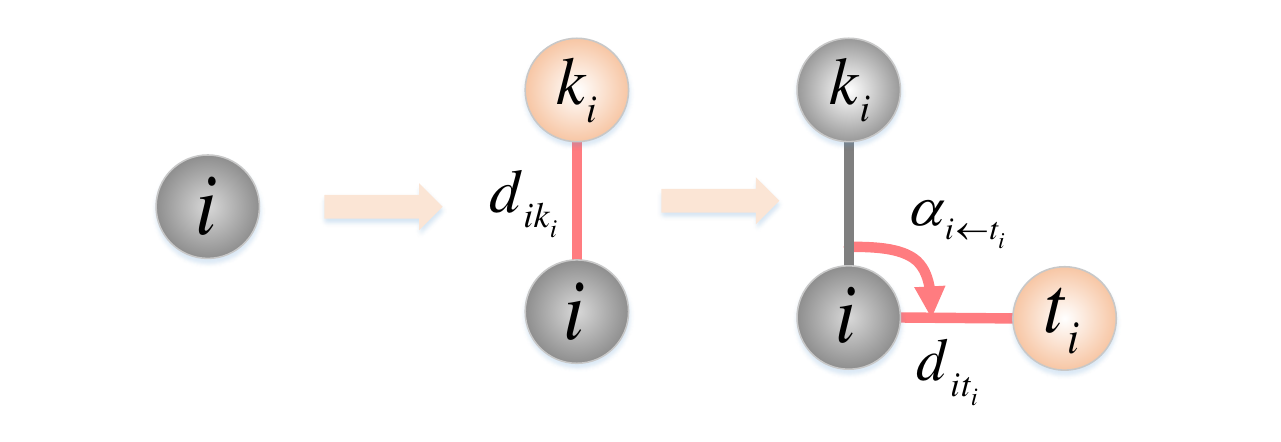}
    \caption{An illustration of the proof process that the geometric \\ structure of $\{atom_i, atom\_k_i, atom\_t_i\}$ is uniquely determined.
    }
    \label{zeroatom}
  \end{subfigure}%
  \begin{subfigure}[b]{0.5\textwidth}
    \includegraphics[clip, trim=2cm 7cm 0 6cm, scale=0.32]{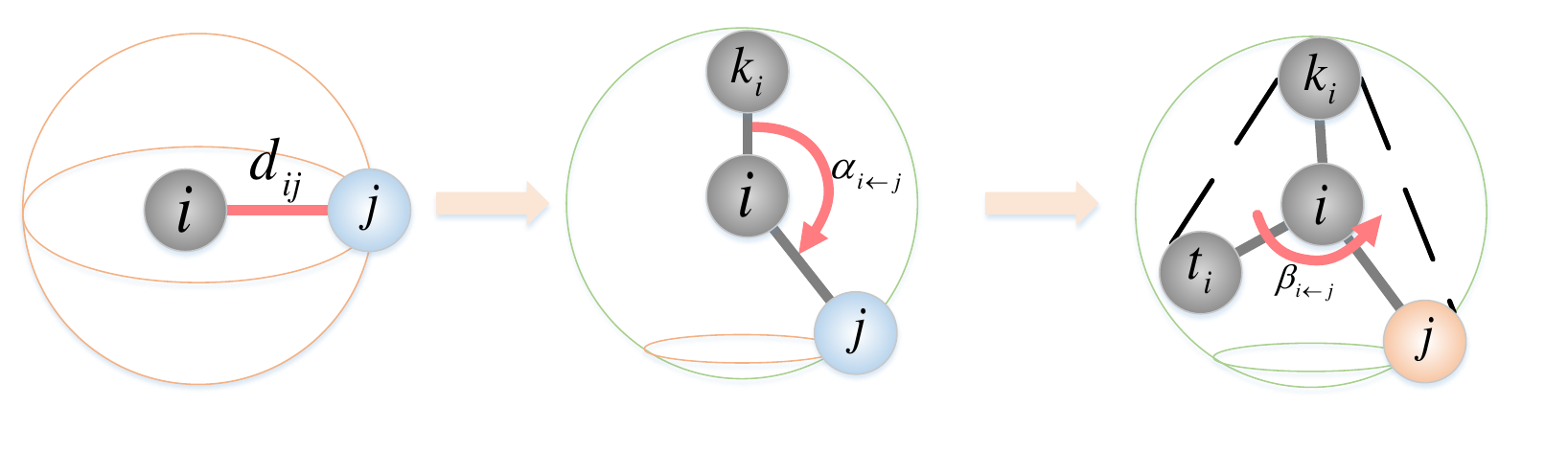}
    \caption{An illustration of the proof process that vector $ \overrightarrow{\mathbf{d_{i j}}}$ is uniquely\\ determined by $\{d_{i j}, \alpha_{i\leftarrow j}, \beta_{i\leftarrow j}\}$.}
    \label{oneatom}
  \end{subfigure}\\[1ex] 
  \begin{subfigure}[b]{0.5\textwidth}
    \includegraphics[clip, trim=2cm 7cm 0 3cm, scale=0.32]{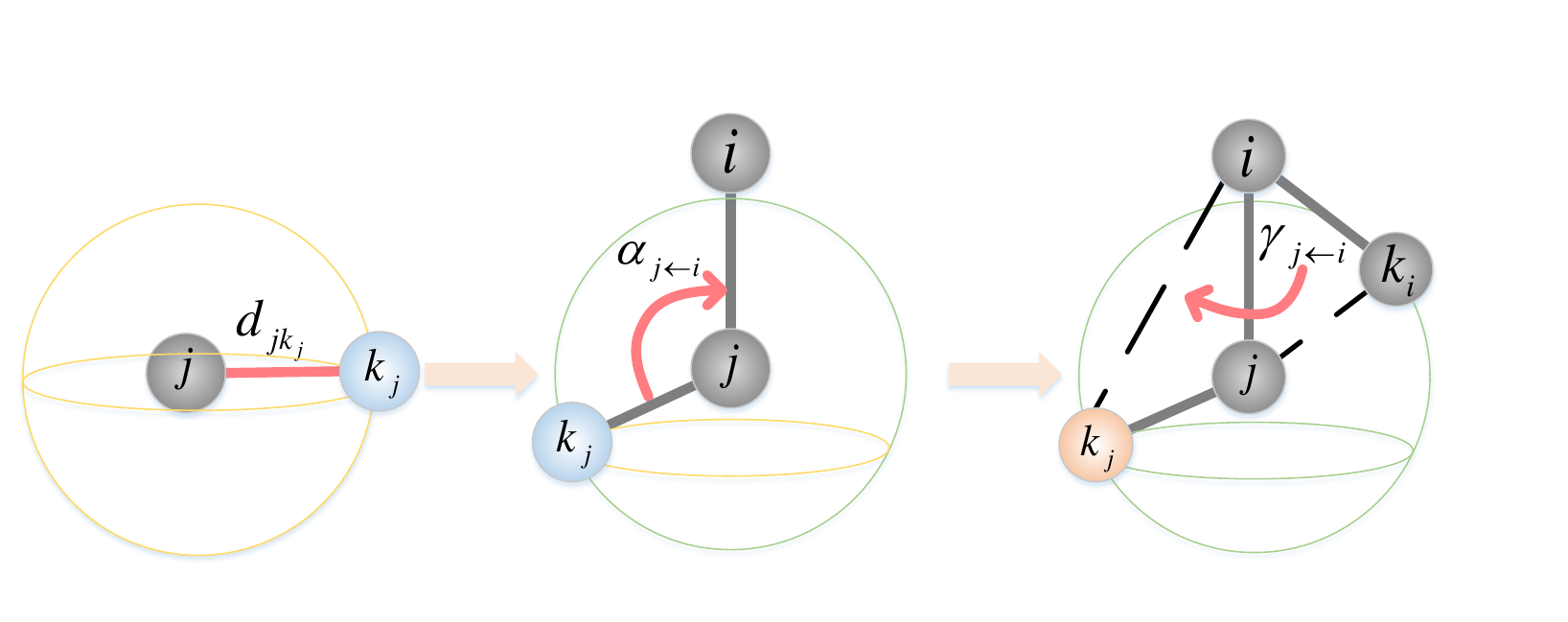}
    \caption{An illustration of the proof process that vector $\overrightarrow{\mathbf{d_{j k_{j}}}}$ is uniquely \\determined by $\{d_{j k_{j}}, \alpha_{j\leftarrow i}, \gamma_{j\leftarrow i}\}$.}
    \label{twoatom}
  \end{subfigure}%
  \begin{subfigure}[b]{0.5\textwidth}
    \includegraphics[clip, trim=2cm 7cm 0 5cm, scale=0.32]{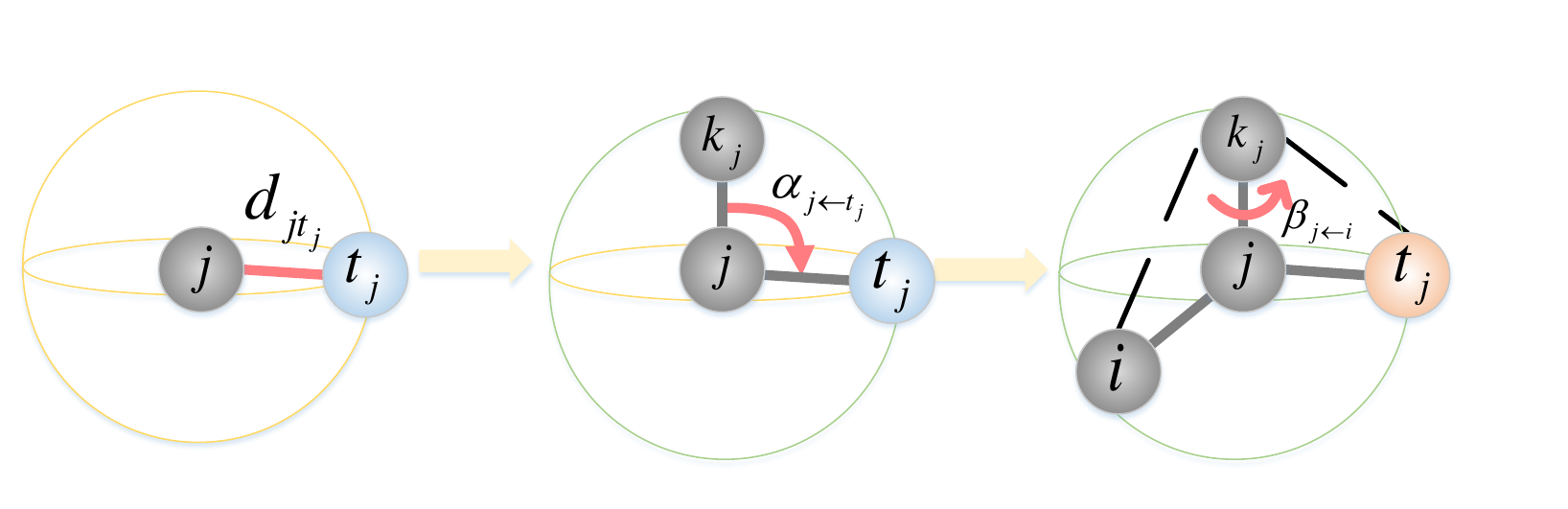}
    \caption{An illustration of the proof process that  vector $\overrightarrow{\mathbf{d_{j t_j}}}$ is uniquely \\determined by $\{d_{j t_j}, \alpha_{j\leftarrow t_j}, \beta_{j\leftarrow i}\}$.}
    \label{threeatom}
  \end{subfigure}
  \caption{Illustrations of the proof process that the geometric structure of cutoff region can be uniquely determined by representation $P=\{d, \alpha, \beta, \gamma\}$.
  }
  \label{fullatom}
\end{figure*}

\begin{theorem}
    For a crystal material $M=(A,X,L)$, its geometric representation  $P_M = [(d_{i j}, \alpha_{i\leftarrow j}, \beta_{i\leftarrow j}, \gamma_{i\leftarrow j})]_{i\in I;j\in NB_i}\in \mathbb{R}^{n_{edge}\times4}$ is periodic complete.
\end{theorem}
Next, we will separately demonstrate the sufficient condition and the necessary condition for THEOREM 3.3, where the definition of periodic completeness is defined in Equation \eqref{complete}.
\begin{proof}
First, let's establish the \textbf{only if ($\Rightarrow$)} condition in Equation  \eqref{complete}. 
This proof will be conducted in two steps.
In the first step, we aim to show that our representation $P$ can uniquely determine the cutoff region, defined as the region composed of atoms within the unit cell and their neighbors. This is achieved by incrementally adding atoms to a uniquely determined geometric structure.

As illustrated in Figure \ref{zeroatom}, we start with only one atom, atom $i$, in the unit cell. Next, we add atom $k_i$, the nearest neighbor of atom $i$. The geometric structure of {atom $i$, atom $k_i$} is uniquely determined by $d_{i k_i}$. Subsequently, we add atom $t_i$, the neighbor of atom $i$, where the angle between vector $\overrightarrow{it_i}$ and vector $\overrightarrow{ik_i}$ is the smallest angle greater than 0 and less than $\pi$. The vector $\overrightarrow{it_i}$ is uniquely determined by $\{d_{i t_i}, \alpha_{i\leftarrow t_i}\}$, so the geometric structure of {atom $i$, atom $k_i$, atom $t_i$} is uniquely determined by $\{d_{i k_i}, d_{i t_i}, \alpha_{i\leftarrow t_i}\}$.


Next, we add atom $j$, a neighbor of atom $i$ in the unit cell. We will show that the vector $\overrightarrow{ij}$ is uniquely determined by $\{d_{i j}, \alpha_{i\leftarrow j}, \beta_{i\leftarrow j}\}$. 
As illustrated in Figure \ref{oneatom}, the position range of atom $j$ is initially a sphere centered on atom $i$ with a radius equal to $d_{i j}$. 
Adding $\alpha_{i\leftarrow j}$ reduces the position range of atom $j$ to a circle. This circle is the intersection of a cone, whose vertex is node i and whose central axis is the line where the vector $\overrightarrow{ik_i}$ is located, with the sphere. The angle between the sides of the cone and the vector $\overrightarrow{ik_i}$ is $\alpha_{i\leftarrow j}$. 
Note that there is a special case where $\alpha_{i\leftarrow j}$ equals 0 or $\pi$. In this case, the position of atom $j$ is the point of intersection of the ray where the vector $\overrightarrow{ik_i}$ is located and the sphere. 
Finally, adding $\beta_{i\leftarrow j}$ uniquely determines the position of atom $j$ to a point, as the intersection point between the new half plane obtained by rotating the half plane $ik_it_i$ along $\beta_{i\leftarrow j}$ and the previously obtained circle. The criteria for selecting atom $t_i$ ensure that atom $i$, atom $k_i$, and atom $t_i$ are not collinear, so half plane $it_ik_i$ bounded by the straight line where vector $\overrightarrow{ik_i}$ located is uniquely determined. 
In summary, vector $\overrightarrow{\mathbf{d_{i j}}}$ is uniquely determined by $\{d_{i j}, \alpha_{i\leftarrow j}, \beta_{i\leftarrow j}\}$.


Next, we add atom $k_j$, the atom $j$'s nearest neighbor except for atom $i$. We will show that the vector $\overrightarrow{jk_{j}}$ is uniquely determined by $\{d_{j k_{j}}, \alpha_{j\leftarrow i}, \gamma_{j\leftarrow i}\}$. As illustrated in Figure \ref{twoatom}, the position range of atom $k_{j}$ is initially a sphere centered on atom $j$ with a radius equal to $d_{jk_{j}}$. Adding $\alpha_{j\leftarrow i}$ reduces the position range of atom $k_{j}$ to a circle. Finally, adding $\gamma_{j\leftarrow i}$ uniquely determines the position of atom $k_{j}$ to a point. In summary, the vector $\overrightarrow{jk_{j}}$ is uniquely determined by $\{d_{j k_{j}}, \alpha_{j\leftarrow i}, \gamma_{j\leftarrow i}\}$.

Next, we add atom $t_j$, the neighbor of atom $j$, where the angle between vector $\overrightarrow{jt_j}$ and vector $\overrightarrow{jk_j}$ is the smallest angle greater than 0 and less than $\pi$. We will show that the vector $\overrightarrow{jt_j}$ is uniquely determined. As illustrated in Figure \ref{threeatom}, the position range of atom $t_j$ is initially a sphere centered on atom $j$ with a radius equal to $d_{j t_j}$. Adding $\alpha_{j\leftarrow t_j}$ reduces the position range of atom $t_j$ to a circle. Finally, adding $\beta_{j\leftarrow i}$ uniquely determines the position of atom $t_j$ to a point. In summary, the vector $\overrightarrow{jt_j}$ is uniquely determined by $\{d_{j t_j}, \alpha_{j\leftarrow t_j}, \beta_{j\leftarrow i}\}$.

Other neighbors of atom $j$ can be added one by one through the same steps shown in Figure \ref{oneatom}. Since the atoms in the crystal material unit cell are strongly connected due to the value of the cutoff, we can repeat the above process until we obtain the entire uniquely determined cutoff region.

Therefore, the cutoff region can be uniquely determined by our representation $P$.

In the next step, we will show that if the cutoff region is uniquely determined, then its corresponding crystal material geometric structure, which consists of $\bm X$ and $\bm L$, is uniquely determined. 
Since the cutoff region is uniquely determined, the unit cell is uniquely determined. 
Since the value of the cutoff is set to be larger than $max\{max\_con, min_{l_1}, min_{l_2}, min_{l_3}\}$, the cutoff region must contain periodic expansions of unit cells along the $\bm l_1$, $\bm l_2$, and $\bm l_3$ directions. Because the cutoff region is uniquely determined, the vector between two atoms with the same index is also determined; that is, the E(3) group of the lattice is also uniquely determined.

Therefore, the \textbf{only if ($\Rightarrow$)} condition in Equation \eqref{complete} holds.
\end{proof}
\begin{proof}
Next, we will prove the \textbf{if ($\Leftarrow$)} condition in Equation \eqref{complete} holds. Since the distances and angles used in our representation $P$ are both relative values that satisfy SE(3) invariance and E(3) invariance, the \textbf{if ($\Leftarrow$)} condition in Equation \ref{complete} holds.
\end{proof}

In summary, we have demonstrated that Equation \eqref{complete} holds based on our geometric representation $P$.

\begin{figure*}[ht]
    \centering
    \includegraphics[clip, trim=3cm 10cm 0 1.5cm, scale=0.78]{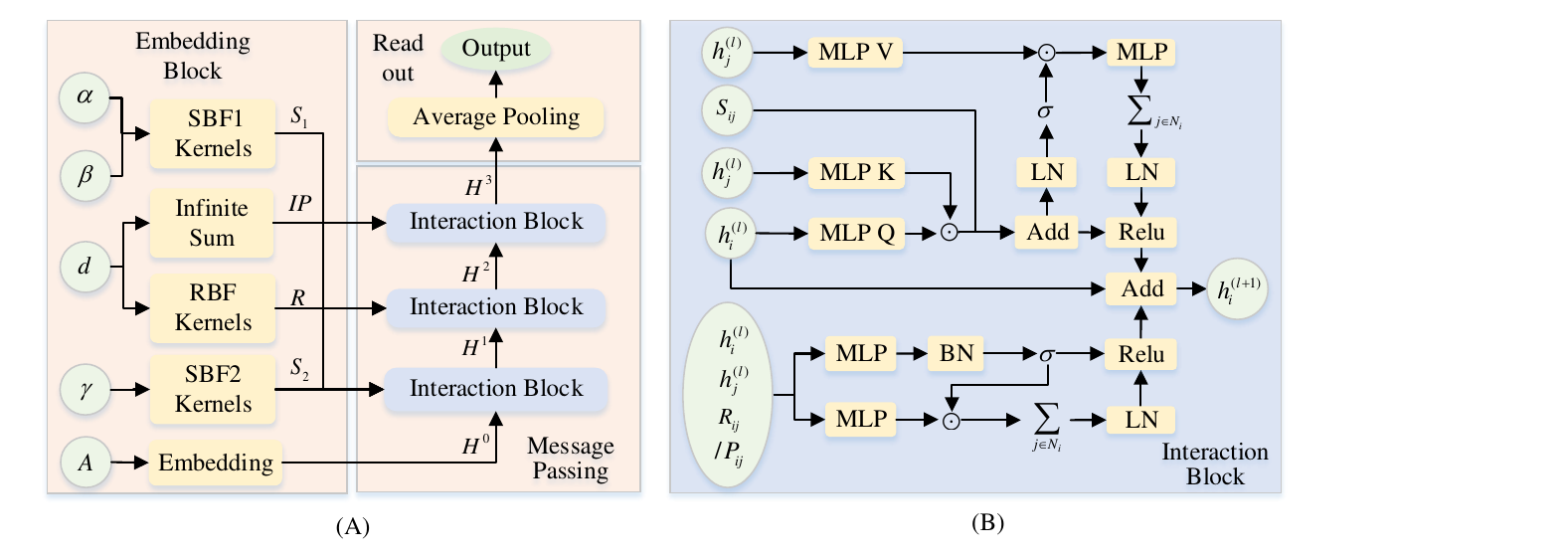}
    \caption{Illustrations of detailed Architecture of PerCNet. 
    (A). An illustration of PerCNet pipeline. 
    (B). An illustration of the detailed interaction block for atom $i$ in (A). 
    The upper half of the network mainly processes angle and dihedral angle information $S$, while the lower half of the network mainly processes distance information, including $R$ and $IP$. 
    }
    \label{all_arc}
\end{figure*}
\section{Proposed Message Passing Scheme}

\subsection{Message Passing Scheme}\label{MessagePassing}

In accordance with the proof of geometric periodic completeness presented in Section \ref{proof}, the periodic complete representation $P$, which satisfies Equation \eqref{complete}, should be constructed based on $(d, \alpha, \beta, \gamma)$ and the cutoff. Consequently, for a crystal material $M$, we have $P_M = \{ d_{i j}, \alpha_{i\leftarrow j}, \beta_{i\leftarrow j}, \gamma_{i\leftarrow j} | \forall i \in I, j \in NB_i\}$, where $I$ represents the atom index set in the unit cell, and $NB_i$ denotes the index set of all neighbors for atom $i$.

Based on the representations that achieve periodic completeness, our model employs a message-passing-based approach to process the input data. The message passing update formula for atom $i$ at layer $l$ is presented as follows:
\begin{equation}
    h_i^{(l)}=g_\phi^{(l)}(h_i^{(l-1)},\sum_{j\in NB_i}f_\theta^{(l)}(h_j^{(l-1)},d_{i j},\alpha_{i\leftarrow j},\beta_{i\leftarrow j}, \gamma_{i\leftarrow j})).
\end{equation}
In this equation, $f_\theta^{(l)}$ and $g_\phi^{(l)}$ represent the neural networks in the $l$-th layer, and $NB_i$ is determined by the cutoff value. The primary goal of the formula is to convolve the features of all neighboring atoms $j$ onto atom $i$. To achieve this, the function $f_\theta^{(l)}$ captures information about the neighbors based on structural properties, while the function $g_\phi^{(l)}$ aggregates the features of all neighboring atoms for atom $i$ to update the features of atom $i$.

\subsection{PerCNet}

Building upon the message-passing scheme described in Section \ref{MessagePassing}, we introduce the \textbf{Per}iodic \textbf{C}omplete \textbf{Net}works (PerCNet), which is depicted in Figure \ref{all_arc}. We adopt a similar framework to the existing crystal representations \cite{lin2023efficient, yan2022periodic, choudhary2021atomistic}, comprising embedding, message passing, and readout modules, as shown on the left side of Figure \ref{all_arc}.

To enhance the aggregation of information from neighboring atoms, we have incorporated several novel components into the message-passing process, such as a transformer-based architecture for weighting the information from bond angles and dihedral angles. In the subsequent sections, we will provide a comprehensive overview of each module.

\subsubsection{Embedding Block}
First, we initialize atom features using the same method as CGCNN \cite{xie2018crystal}, where the initial value of atom features is the embedding of the atom number and its corresponding group.
Next, for the geometric structures of crystal molecules, we utilize two relative quantities, distance, and angle, to represent them. It's important to note that we did not directly use $P_M$ as inputs to the network. Instead, we transformed them into more physically meaningful variables. Given the significance of distance, we extended it on the sphere using the previously mentioned Radial Basis Function (RBF). Additionally, we further processed it into the $IS$ variable using the infinite summation method employed in Potnet \cite{lin2023efficient}. For angles, we extended them using Spherical Basis Function (SBF) to convert them into physically meaningful variables. 
Spherical Basis Function (SBF) is a method for representing three-dimensional spatial data in a spherical coordinate system. These physically meaningful vectors, RBF and SBF, are commonly used to represent the geometric structures of molecules and materials in other works such as GemNet \cite{gasteiger2021gemnet}, SphereNet \cite{liu2021spherical}, ComENet \cite{wang2022comenet}, DimeNet \cite{gasteiger2020directional}, Matformer \cite{yan2022periodic}, and PotNet \cite{lin2023efficient}.

The formulas for RBF and SBF are $
    RBF(d)=j_l(\frac{\theta_{ln}}{c}d),
    SBF_1(\gamma)=Y_l^m(\gamma),
    SBF_2(\alpha,\beta)=Y_l^m(\alpha, \beta)$.
Here, $j_l$ is a spherical Bessel function of order $l$, and $\theta_{ln}$ is the $n$-th root of the $l$-order Bessel function. The cutoff is represented by $c$. $Y_l^m$ is a spherical harmonic function of degree $m$ and order $l$.

\subsubsection{Message Passing Block and Readout Block}
The procedure for updating the feature embedding $h_i^{(l)}$ of atom $i$ is depicted on the right side of Figure \ref{all_arc}. Specifically, we employ a two-part network to handle the two physically meaningful vectors derived from $RBF$ and $SBF$. On one hand, we apply an MLP-based convolution to process the distance embedding. On the other hand, we utilize a transformer-based convolution for the angle and dihedral angle embeddings, denoted as $S_1$ and $S_2$. Here, MLP Q, MLP K, and MLP V denote multilayer perceptrons, while $\circ$ and $\sigma(x)$ represent the Hadamard product and sigmoid function, respectively. The resulting value is then added to the initial $h_i^{(l)}$ to generate the updated atom feature $h_i^{(l+1)}$.
Lastly, the readout block utilizes average pooling to compile and consolidate the atom features from all atoms within the unit cell. The purpose of this block is to produce the final predictions.
\section{Experiments}

\subsection{Experimental Setup}
We evaluate the effectiveness and periodic completeness of PerCNet on two large-scale material benchmark datasets, The Materials Project-2018.6 and JARVIS, as well as one dataset with similar crystal structures. The baseline methods include CGCNN \cite{xie2018crystal}, SchNet \cite{schutt2017schnet}, MEGNET \cite{chen2019graph}, GATGNN \cite{louis2020graph}, ALIGNN \cite{choudhary2021atomistic}, Matformer \cite{yan2022periodic}, and PotNet \cite{lin2023efficient}.

For all tasks, we utilize one NVIDIA GeForce RTX 24G 3090 GPU for computation. In terms of implementation, all the models are trained using the Adam optimizer \cite{kingma2014adam} with a one-cycle learning rate scheduler \cite{smith2019super}, featuring a learning rate of 0.001, a batch size of 64, and a training epoch count of 500.

\subsection{Experimental Result}

\begin{table}[t]
  \centering  
  \begin{threeparttable}  
  \caption{Statistics of datasets.}  
  \label{tab:Dataset} 
    \begin{tabular}{c|ccc}
    \toprule  
    Tasks&\# training&\# validation&\# testing\cr
    \midrule  
    Formation Energy (JV)&44578&5572&5572\cr
    Total energy (JV)&44578&5572&5572\cr
    Band Gap MBJ (JV)&14537&1817&1817 \cr
    Ehull (JV)&44296&5537&5537\cr
    Formation Energy (MP)&60000&5000&4239\cr
    Band Gap (MP)&60000&5000&4239\cr 
    \bottomrule  
    \end{tabular}  
    \end{threeparttable}  
\end{table}
First, we evaluate our model on two widely used large-scale material benchmark datasets: The Materials Project-2018.6.1 dataset \cite{jain2013materials} with 69,239 crystals and the JARVIS-DFT-2021.8.18 3D dataset \cite{choudhary2020joint} with 55,722 crystals. We noticed that prior works \cite{xie2018crystal,choudhary2021atomistic,schutt2017schnet} used different proportions or random seeds for dataset partitioning. To ensure a fair comparison, we adopt the same data settings as previous works \cite{choudhary2021atomistic,yan2022periodic,lin2023efficient} across all tasks for all baseline models and report the number of data in Table \ref{tab:Dataset}.
\begin{table*}[t]  
  \centering  
  \begin{threeparttable}  
  \caption{Comparison between our method and other baselines in terms of test MAE on JARVIS dataset and The Materials Project dataset. The best results are shown in {\bf bold} and the second best results are shown with \underline {underlines}.}  
  \label{tab:Jarvis and MP} 
    \begin{tabular}{c|cccccc}
    \toprule  
    \multirow{3}{*}{Method}&
    \multicolumn{4}{c|}{JARVIS dataset }&\multicolumn{2}{c}{Materials Project dataset}\cr
    \cmidrule(lr){2-7}
    &\multicolumn{1}{c}{Formation Energy}&\multicolumn{1}{c}{Total energy}&\multicolumn{1}{c}{Bandgap(MBJ)}&\multicolumn{1}{c}{Ehull}&\multicolumn{1}{c}{Formation Energy}&\multicolumn{1}{c}{Band Gap}\cr  
    \cmidrule(lr){2-7}     
    &eV/atom&eV/atom&eV&eV&eV/atom&eV\cr  
    \midrule  
    CGCNN&0.063&0.078&0.41&0.17&0.031&0.292\cr
    SchNet&0.045&0.047&0.43&0.14&0.033&0.345\cr
    MEGNET&0.047&0.058&0.34&0.084&0.030&0.307\cr
    GATGNN&0.047&0.056&0.51&0.12&0.033&0.280\cr
    ALIGNN&0.0331&0.037&0.31&0.076&0.0221&0.218\cr
    Matformer&0.0325&0.035&0.30&0.064&0.0210&0.211\cr
    PotNet&\underline {0.0294}&\underline {0.032}&\underline {0.27}&\underline {0.055}&\underline {0.0188}&\underline {0.204}\cr
    Ours&\bf 0.0287&\bf 0.0307&\bf 0.2657&\bf 0.0503&\bf 0.0181&\bf 0.2007\cr  
    \bottomrule  
    \end{tabular}  
    \end{threeparttable}  
\end{table*}

The evaluation metric used is the test Mean Absolute Error (MAE), following previous studies \cite{choudhary2021atomistic, lin2023efficient, yan2022periodic, xie2018crystal}. As seen in Table \ref{tab:Jarvis and MP}, PerCNet achieves the best performance across all tasks on both benchmark datasets. Specifically, it reduces MAE by $1.59\%\sim8.54\%$ compared to the previous state-of-the-art (SOTA) PotNet \cite{lin2023efficient} for all tasks, which is a significant improvement. In summary, these superior performances demonstrate the effectiveness of our periodic complete crystal representation method.

\begin{table}[h]  
  \centering  
  \begin{threeparttable}  
  \caption{Total Training Time and Total Testing Time compared with ALIGNN and PotNet on JARVIS formation energy prediction.
  } 
  \label{tab:Training time} 
    \begin{tabular}{cccc}
    \toprule  
    Method&Training Time&Testing Time&Per Testing Crystal\cr   
    \midrule  
    ALIGNN&37.83h&5.41min&61ms\cr   
    Potnet&8.037h&1.081min&12ms\cr  
    Ours&19.367h& 1.92min&20ms\cr  
    \bottomrule  
    \end{tabular}  
    \end{threeparttable}  
\end{table}

We also analyze the time cost of the periodic complete algorithm in Table \ref{tab:Training time}. We compare the training time and testing time using the JARVIS formation energy dataset with the most related method, ALIGNN, which also computes angles, and the most recent method, PotNet. 
Compared with ALIGNN, our PerCNet is more efficient. As shown in Table \ref{tab:Training time}, our PerCNet is nearly two times faster than ALIGNN in terms of total training time and three times faster in inference time for the whole test set. 
Although PerCNet is slower than PotNet, as PotNet only computes distance, it is much more effective.
\subsection{Periodic Completeness Experiment}

\begin{figure}
  \centering
  \begin{minipage}[t]{0.45\textwidth} 
    \centering
    \includegraphics[clip, trim=0 0 0 5cm,scale=0.23]{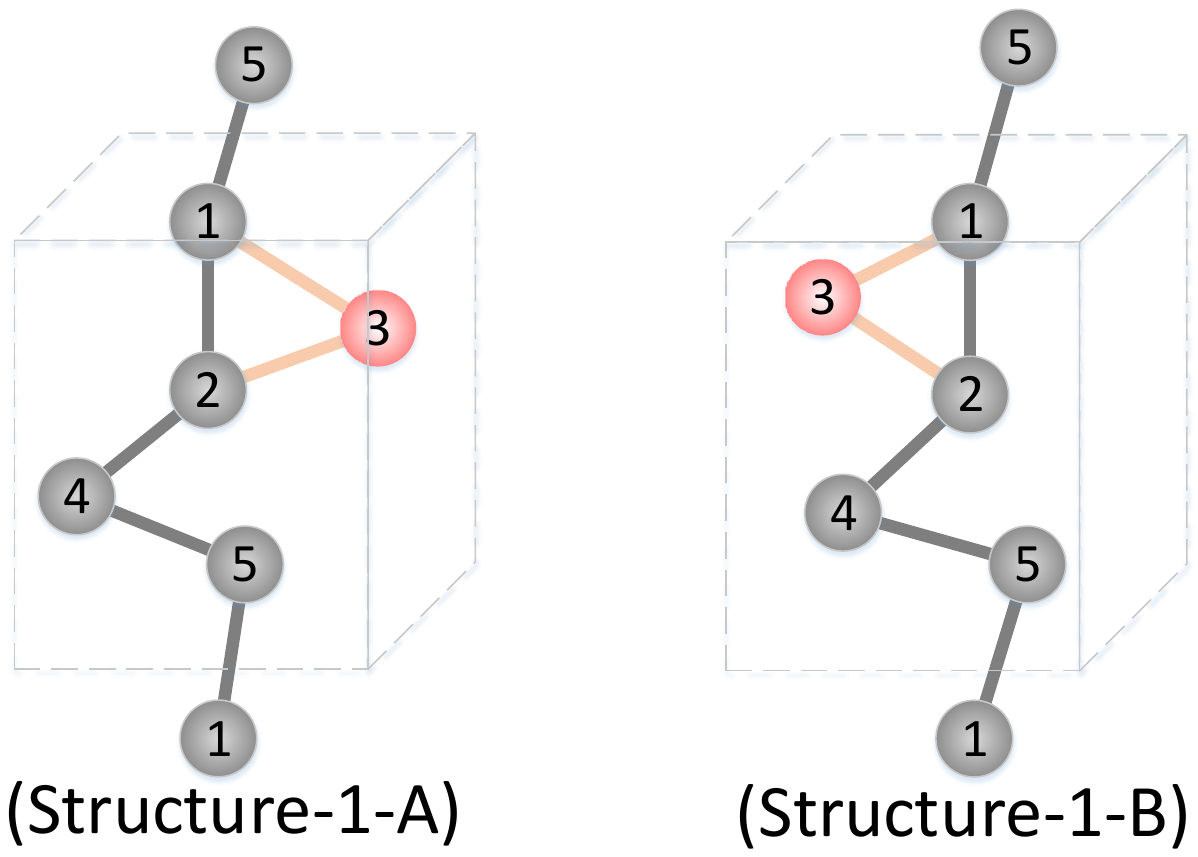}
  \end{minipage}
  \begin{minipage}[b]{0.45\textwidth} 
    \centering
    \includegraphics[clip, trim=0 5cm 0 0,scale=0.23]{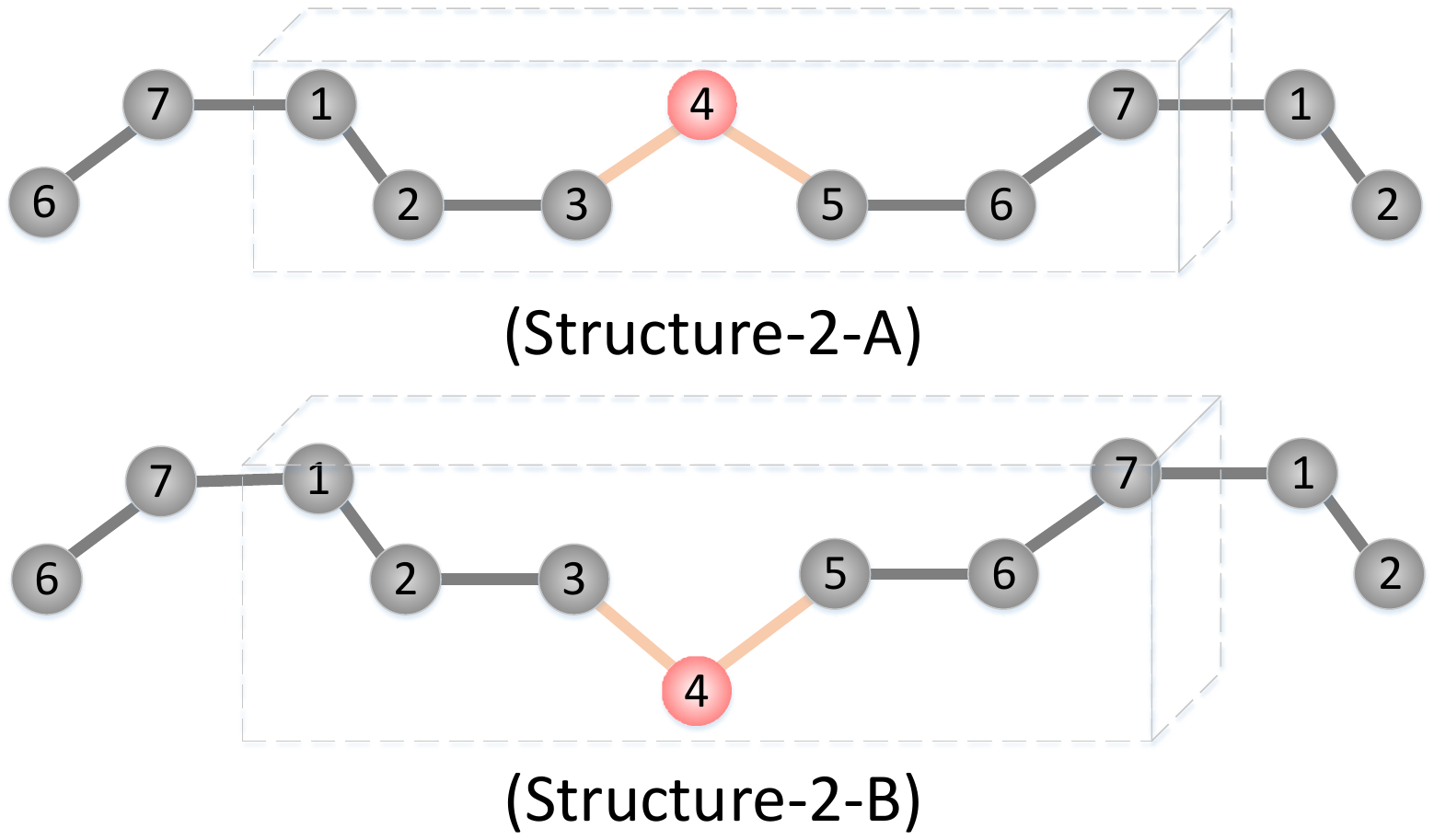}
  \end{minipage}
  \caption{Illustrations of two kinds of similar periodic structures that other crystal representation methods can not distinguish due to lack of periodic completeness. (Structure-1-A). An illustration of a crystal molecule with five atoms within the unit cell. (Structure-1-B). An illustration of a crystal molecule that has a similar structure to the molecule illustrated in Structure-1-A, except that the position of atom 3 has changed, while the distances between atom 3 and atom 2 and between atom 3 and atom 4 remain unchanged. (Structure-2-A). An illustration of a crystal molecule with seven atoms within the unit cell. (Structure-2-B). An illustration of a crystal molecule that has a similar structure to the molecule illustrated in Structure-2-A, except that the position of atom 4 has changed, while the distances between atom 4 and atom 3 and between atom 4 and atom 5 remain unchanged.
  }
  \label{fig:Structure}
\end{figure}

Given the near-infinite number of crystal materials, it's impractical to experimentally prove that a representation can distinguish all crystal structures. Therefore, we use visualized structures depicted in Figure \ref{fig:Structure} to show that our work can provide better representations for crystal graphs compared to others.

 As shown in Table \ref{tab:completeness}, PerCNet successfully identified differences between these structures, whereas other works considered the structures to be the same. This superior performance of PerCNet showcases the effectiveness of our periodic complete representation.

\begin{table}[t]  
  \centering  
  \caption{The comparison among our method and CGCNN, PotNet in terms of representation similarity on two kinds of similar but different crystal structures. 
  }  
  \label{tab:completeness} 
    \begin{tabular}{ccc}
    \toprule  
    \multirow{2}{*}{Method}
    &\multicolumn{1}{c}{Structure-1} & \multicolumn{1}{c}{Structure-2}\cr
    \cmidrule(lr){2-3}
    &Similarity$\downarrow$&Similarity$\downarrow$ \cr  
    \midrule  
    CGCNN&100\%&100\%\cr  
    PotNet&100\%&100\%\cr
    Ours& \bf{92.857\%}&\bf{90\%}\cr  
    \bottomrule  
    \end{tabular}  
\end{table}

\subsection{Ablation Studies}
\begin{table}[ht]  
  \centering  
  \begin{threeparttable}  
  \caption{Ablation studies for the effects of ignoring periodicity or completeness.} 
  \label{tab:Ablation studies} 
    \begin{tabular}{cc}
    \toprule  
    \multirow{2}{*}{Method}&
    \multicolumn{1}{c}{JARVIS Formation Energy}\cr  
    \cmidrule(lr){2-2}     
    &eV/atom\cr  
    \midrule  
    Ours&\bf{0.0287}\cr  
    Ours without periodicity&0.0357 ($24.39\%\downarrow$) \cr
    Ours without Completeness&0.0294 ($2.44\%\downarrow$) \cr  
    \bottomrule  
    \end{tabular}  
    \end{threeparttable}  
\end{table}
In this section, we showcase the significance of two core components of PerCNet, which are the periodic scheme and cutoff region completeness, for crystal prediction. We conduct experiments on the JARVIS 
formation energy task and use test MAE for evaluation.

\subsubsection{Periodic scheme}
We highlight the importance of the periodic scheme by contrasting the full PerCNet model with `Ours without periodicity'. The latter focuses only on the unit cell and disregards the periodicity of the crystal structure. As evidenced in Table \ref{tab:Ablation studies}, the absence of the periodic scheme resulted in a $24.39\%$ drop in performance, underscoring the significance of periodic representation.

\subsubsection{Completeness}
The importance of unit cell completeness is shown by comparing the PerCNet model with `Ours without completeness', which solely models distance and omits the role of global information, including angles. As depicted in Table \ref{tab:Ablation studies}, the lack of angle information in the representation prevents the algorithm from capturing the complete crystal structure, leading to a performance drop from 0.0287 to 0.0294 in the formation energy prediction task.
\section{Conclusion}

In this paper, we target the many-to-one issue in crystal representation.
We first propose a periodic complete representation for crystal material graphs by considering dihedral angles of first-order neighbors. 
A theoretical proof for the periodic completeness of the proposed representation is provided, which can guarantee the proposed representation corresponds one-to-one with the crystal material. Based on the proposed representation, we then propose PerCNet with a specially designed message passing mechanism for crystal property prediction tasks. Experimental results on two large-scale datasets including The Materials Project and JARVIS show that PerCNet outperforms baseline methods including PotNet (SOTA). In addition, ablation experimental results demonstrate the importance of periodic schemes and completeness. 
In summary, we provide a powerful representation algorithm and architecture for the field of crystal molecules. In the future, we expect to apply this representation method to more fields of crystal materials, such as crystal structure generation.

\bibliographystyle{unsrt}  
\bibliography{references}

\end{document}